# Tkachenko modes and quantum melting of Josephson junction type of vortex array in rotating Bose Einstein condensate

Aranya B Bhattacherjee*

INFM, Dipartimento di Fisica E.Fermi, Universita di Pisa, Via Buonarroti 2, I-56127, Pisa, Italy

**Abstract:** Using path integral formalism, we show that the Abrikosov-Tkachenko vortex lattice may equivalently be understood as an array of Josephson junctions. The Tkachenko modes are found to be basically equivalent to the low energy excitations (Goldstone modes) of an ordered state. The calculated frequencies are in very good agreement with recent experimental data. Calculations of the fluctuations of the relative displacements of the vortices show that vortex melting (quantum phase transition) is a result of quantum fluctuations around the ordered state due to the low energy excitations (Tkachenko modes) and occurs when the ratio of the kinetic energy to the potential energy of the vortex lattice, $E_C/E_J \geq 0.001$.

*Permanent Institute: Department of Physics, Atma Ram Sanatan Dharma College, University of Delhi (South campus), Dhaula Kuan, New Delhi-110021, India.

Vortices in a superfluid formed due to rapid rotation tend to organize into a triangular lattice (also known as Abrikosov lattice) as a result of mutual repulsion. Interestingly, the Abrikosov lattice has an associated rigidity. In the recent years, it has become possible to achieve such a vortex lattice state in a rapidly rotating Bose-Einstein condensate [1-4]. In 1966, Tkachenko proposed that a vortex lattice in a superfluid would support transverse elastic modes [5]. The possibility of attaining Tkachenko oscillations in BEC was proposed recently [6] and observations of these modes in Bose condensed [87]Rb were first reported by Coddington et.al. [7] at rotation speeds ($\Omega$) up to 0.975 of the transverse trapping frequency $\omega_\perp$. Subsequent theoretical models were put forward to explain the experimental results [8]. These new developments both in theory and experiments have opened a new direction for the study of quantum vortex matter, such as fractional quantum Hall like and melted vortex lattice states [9,10]. This new state of quantum vortex matter occurs only at very rapid rotations. When $\Omega$ exceeds a certain critical value (determined by the system parameters), the system enters the "mean field" quantum Hall regime, in which the condensation is only in lowest Landau level orbits. Eventually the vortex lattice melts, and the system enters a strongly correlated regime.

Motivated by these interesting developments, in this work, we extend the ideas that were developed in connection with "Josephson junction arrays" in superconductors to an array of vortices in BEC. We name this system as "Josephson junction vortex array (JJVA)". Our starting point will be the hydrodynamic vortex Hamiltonian [9,10]. The present analysis will be restricted to two dimensions transverse to the rotation axis. We calculate the Tkachenko modes using the path integral formalism applied to the JJVA. We will be basically working in the regime where the inter-lattice spacings are constant at a given $\Omega$.

We consider a large number of vortices in a rotating gas of bosons at temperatures well below the Kosterlitz-Thouless transition temperature given by $T_c^{KT} = \hbar^2 \rho_S / 4m$, where $\rho_S$ is the particle density. The BEC is described by a repulsive short range interaction $U(\vec{r}) = g\delta^3(\vec{r})$, with $g = 4\pi\hbar^2 a_s / m$, where $a_s$ is the s-wave scattering length. The superfluid forms a triangular lattice of quantized vortices (carrying the angular momentum of the system) rotating as a solid body at angular velocity $\Omega$. We assume that at low temperatures, the scattering of thermal excitations by vortices does not affect the vortex dynamics. The situation considered then corresponds to considering a vortex as a point particle moving under the influence of the Magnus force. The vortex motion is governed by a Hamiltonian corresponding to one of point particles, with a charge equal to the quantum of circulation $\kappa = h/m$, interacting with electromagnetic fields. For a large collection of vortices $N_V \gg 1$ in a frame rotating with angular velocity $\Omega$, the vortex Hamiltonian reads:

$$H_V = \sum_{i=1}^{N_V} \frac{(\vec{P}_i - \kappa \vec{A}_i)^2}{2m_v} - \Omega(\vec{X}_i \times \vec{P}_i)_z - \frac{\rho_S \kappa^2}{2\pi} \sum_{i<j}^{N_V} \ln\left|\frac{\vec{X}_i - \vec{X}_j}{\xi}\right| \tag{1}$$

The first term in Eqn.(1) is the kinetic energy of the vortices, the second term is a result of centrifugal and coriolis forces on vortices and the last term is the repulsive logarithmic Coulomb interaction of point vortices. $\xi$ is the coherence length and is $\xi = \sqrt{\hbar/2g\rho_S}$. The effective vortex mass $m_V = \pi \rho_S \xi^2$ [10] is very small since the coherence length is small. Hence, we will take below the limit $m_V \to 0$. The pseudo vector potential due to the Magnus force is $A_i^a = \frac{1}{2}\rho_S \varepsilon^{ab} X_b^i$ ($a,b = x,y$) [10]. The canonical momentum of the vortex $i$ is given as:

$$\vec{P}_i = m_V(\dot{\vec{X}}_i + \vec{\Omega} \times \vec{X}_i) + \kappa \vec{A}_i \tag{2}$$

The Hamiltonian of Eqn. (1) is rewritten in the small vortex mass limit as:

$$H_V = \frac{1}{2}\rho_S \kappa \Omega \sum_{i=1}^{N_V} \vec{X}_i^2 - \frac{\rho_S \kappa^2}{2\pi} \sum_{i<j}^{N_V} \ln\left|\frac{\vec{X}_i - \vec{X}_j}{\xi}\right| \tag{3}$$

The ground state of the system is given by the spatially homogeneous vortex density $n_V = 2\Omega/\kappa$. To develop the main aim of this work in the simplest way, we consider a lattice of singly quantized vortices with lattice spacing $d$, each of which is coupled to its nearest neighbors by a link with the Josephson coupling energy $E_J = \rho_S \kappa^2 / 2\pi$. If we focus our attention on one vortex, then the nearest neighbors tends to screen it from the other vortices and hence the interaction becomes short range. Further the periodicity of the interaction is maintained i.e. if we displace one vortex by an amount $\Delta d$, then all vortices are displaced by a similar amount due to mutual repulsion and $(\vec{X}_i - \vec{X}_j)/d = \hat{\tilde{X}}_i - \hat{\tilde{X}}_j$ remains unchanged. This physical picture corresponds to the "stiff" regime where $k \approx 1/R$, is the lowest wavenumber in the finite geometry and $R$ is the size of the system transverse to the rotation axis [8]. In the stiff regime the rotational frequency $\Omega < sk$, where $s$ is the velocity of sound and the condensate behaves effectively as an incompressible fluid. At higher rotational frequencies, the compressibility plays an important role in response to rotation. Consequently, to a good approximation, we can replace the repulsive logarithmic coulomb interaction by a periodic one i.e. $\ln\left|\frac{\vec{X}_i - \vec{X}_j}{\xi}\right| \approx \left\{1 - \cos\left(\hat{\tilde{X}}_i - \hat{\tilde{X}}_j\right)\right\}$, valid in the region $\xi \leq (\vec{X}_i - \vec{X}_j) \leq d$. The Hamiltonian is now invariant under the transformation $(\hat{\tilde{X}}_i - \hat{\tilde{X}}_j) \to (\hat{\tilde{X}}_i - \hat{\tilde{X}}_j) + 2\pi$. The above physical picture of the vortex array allows us to rewrite the Hamiltonian of eqn. (3) as:

$$H_V = \frac{1}{2}\rho_S \kappa \Omega d^2 \sum_{i=1}^{N_V} \hat{\tilde{X}}_i^2 - \frac{\rho_S \kappa^2}{2\pi} \sum_{<ij>}^{N_V} \left\{1 - \cos\left(\hat{\tilde{X}}_i - \hat{\tilde{X}}_j\right)\right\} \qquad (4)$$

Where $<ij>$ runs over all neighboring lattice points. Using the relation between the momentum and the displacement of the $i^{th}$ vortex $P_i = -i\hbar \frac{\partial}{\partial X_i}$, we rewrite the Hamiltonian of eqn.(4) as:

$$H_V = -\frac{2\Omega\hbar^2}{d^2 \rho_S \kappa} \sum_{i=1}^{N_V} \left(\frac{\partial}{\partial \hat{\tilde{X}}_i}\right)^2 - \frac{\rho_S \kappa^2}{2\pi} \sum_{<ij>}^{N_V} \left\{1 - \cos\left(\hat{\tilde{X}}_i - \hat{\tilde{X}}_j\right)\right\} \qquad (5)$$

The Hamiltonian of eqn.(5) is similar to the Hamiltonian describing the quantum dynamics of an array of Josephson junction (JJ) [11]. Consequently, the dynamics of the vortex array is similar to that of Josephson junctions and hence we refer it to as JJVA. In the original JJ model a quantum phase transition (QPT) from a superconducting-insulator state is driven by the competition between two physical magnitudes: the Josephson coupling energy and $E_J$, which governs the tunneling through the intrawell barriers, and the on-site interparticle interaction energy $E_C$. When the ratio $E_C/E_J < 1$, the system is a superconductor, and when $E_C/E_J > 1$, the system becomes an insulator. Equivalently, $E_J$ corresponds to the potential energy of the system and $E_J$ represents the kinetic energy. Analogous to the JJ model, we identify in our model from eqn.(5), $E_C/E_J = 0.28/v^2$ ($v = N/N_V$ is the filling fraction). A QPT, from an ordered vortex lattice to a melted vortex lattice state is then expected for $v \leq v_c$, where $v_c$ is some critical value of the filling fraction at which the lattice is expected to melt. We defer our discussion of vortex lattice melting for later.

Quantization of the Hamiltonian of eqn.(5) with path integral methods leads to the partition function:

$$Z = \int \widetilde{D}\hat{\vec{X}}_i(\tau) \exp\left[-\int_0^{\beta\hbar} d\tau \left\{ \frac{\rho_S \kappa d^2}{8\hbar\Omega} \sum_i \left(\dot{\hat{\vec{X}}}_i(\tau)\right)^2 + \frac{\rho_S \kappa^2}{2\pi\hbar} \sum_{<ij>} \left\{1 - \cos\left(\hat{\vec{X}}_i - \hat{\vec{X}}_j\right)\right\}\right\}\right] \quad (6)$$

Where $\widetilde{D}\hat{\vec{X}}_i(\tau)$ is multiple integrals over $\hat{\vec{X}}_i$ and $\beta = 1/k_B T$. Consider now a small variation $\delta\hat{\vec{X}}_i$ in the displacement $\hat{\vec{X}}_i$. We expand around this point the action up to second order in the variation $\delta\hat{\vec{X}}_i$.

$$S_0 = \int_0^{\beta\hbar} d\tau \left[ \frac{\rho_S \kappa d^2}{8\hbar\Omega} \sum_i \left(\delta\dot{\hat{\vec{X}}}_i(\tau)\right)^2 + \frac{\rho_S \kappa^2}{4\pi\hbar} \sum_{<ij>} \left[\delta\hat{\vec{X}}_i(\tau) - \delta\hat{\vec{X}}_j(\tau)\right]^2 \right] \quad (7)$$

Performing a Fourier transformation

$$\delta\hat{\vec{X}}_i = \frac{1}{\sqrt{N_V \beta}} \sum_n \exp(-i\omega_n \tau) \exp(i\vec{k}.\vec{X}_i) X(i\omega_n, \vec{k}) \quad (8)$$

the action becomes:

$$S_0 = \sum_n \sum_{\vec{k}} \left[ \frac{\rho_S \kappa d^2}{8\hbar\Omega} \omega_n^2 + \frac{\rho_S \kappa^2}{2\pi\hbar} \sum_{l=1}^{D} (1 - \cos(k_l d)) \right] \hat{\vec{X}}(i\omega_n, \vec{k}) \hat{\vec{X}}(-i\omega_n, -\vec{k}) \quad (9)$$

$D$ is the dimension of the system. Comparing eqn.(9) with the partition function of an ensemble of harmonic oscillators:

$$Z = \int \prod_n dx_n \exp\left[-\sum_n \frac{m}{2}\left(\omega_n^2 + \omega^2\right)|x_n|^2\right] \qquad (10)$$

We obtain the frequency of the vortex lattice as:

$$\omega_T^2 = \frac{8\hbar\Omega}{md^2} \sum_{l=1}^{D} (1 - \cos k_l d) \qquad (11)$$

Where the lattice spacing $d = 1.9\sqrt{\hbar/m\Omega}$. In particular, for small $|\vec{k}|$, $\cos k_l d$ can be expanded, and as a result

$$\omega_T^2 = \frac{4\hbar\Omega}{m}|\vec{k}|^2 \qquad (12)$$

and we obtain Tkachenko waves with velocity $\left(\frac{4\hbar\Omega}{m}\right)^{1/2}$. To compare eqn.(12) with the experiment of ref.(7), we must have the effective value of $k = \alpha/R$. $R$ is the Thomas-Fermi radius in the transverse direction [8,12]

$$R^2 = a_\perp^2 \gamma (1-x^2)^{-3/5} \qquad (13)$$

Where $x = \Omega/\omega_\rho$, $\gamma = \left[\left(15N\frac{a_s}{a_\rho}\right)\left(\frac{\omega_z}{\omega_\rho}\right)\right]^{2/5}$ and $a_\rho = \sqrt{\hbar/m\omega_\rho}$ is the transverse oscillator length, $\omega_z$ is the axial trapping frequency. $\alpha$ is a constant which is found from the best fit of eqn.(12) to the experimental data. This gives $\alpha = 1$ for $N = 2.5 \times 10^6$, $\{\omega_z, \omega_\rho\} = 2\pi\{8.3, 5.2\}$Hz. Fig.1 shows the Tkachenko mode frequencies over the entire range of $\Omega$ together with the experimental points of ref.[7].

Let us now calculate the fluctuations in the relative displacement of $i^{th}$ and the $j^{th}$ vortex $\left(<\left(\delta\hat{\vec{X}}_i - \delta\hat{\vec{X}}_j\right)^2>\right)$. This is useful to estimate the point where the lattice is expected to melt at T=0. With eqns.(8) and (9), we can obtain for a 2-D pancake like BEC

$$<\left(\delta\hat{\vec{X}}_i - \delta\hat{\vec{X}}_j\right)^2> = \int \frac{d^2\vec{k}}{(2\pi)^2} \frac{2d^2 E_C}{\omega_T \hbar}\left[n_B(\omega_T) + \frac{1}{2}\right](1 - \cos\vec{k}d) \qquad (14)$$

Where $E_C = 4\hbar^2\Omega/\rho_s \kappa d^2$ and the Bose distribution function $n_B(\omega_T) \approx k_B T/\hbar\omega_T$ for $\hbar\omega_T \ll k_B T$.

In the low frequency limit, we can replace [11] $\int \frac{d^2\vec{k}}{(2\pi)^2} \to n_V$ and $(1-\cos\vec{k}d) \approx \frac{1}{2}\vec{k}^2 d^2$.

We then have for T=0

$$\frac{<(\delta\hat{\vec{X}}_i - \delta\hat{\vec{X}}_j)^2>}{R_{ws}^2} \approx \frac{1.43d}{\nu R} \approx \frac{3.17(1-x^2)^{3/10}}{\nu\sqrt{x}} \tag{15}$$

Where $R_{ws}^2 \approx 0.28d^2$ is the radius of the Wigner Seitz cell around a given vortex [12]. $\nu = N/N_V$ is the filling fraction. Taking the Lindemann criterion for melting of a vortex lattice in 2-D put forth in ref.[14], $<(\delta\hat{\vec{X}}_i - \delta\hat{\vec{X}}_j)^2>/R_{ws}^2 \approx 0.07$, we plot the filling fraction $\nu$ as a function of $\Omega/\omega_\rho$ in fig.2. The dephasing of the condensate due to vortex melting becomes significant only when $N_V \to N$. This occurs only at rapid rotations. For $\Omega = 0.975\omega_\rho$ (maximum rotation achieve until now [7], we find from fig.2, melting of the vortex lattice is expected to occur at $\nu \approx 17$ (in accordance with ref.[12]). With this value of filling fraction, we find that the ratio of the kinetic energy to the potential energy which determines the QPT is $E_C/E_J = 0.001$. Owing to the commutation relation $[X_i, P_j] = i\hbar\delta_{ij}$, the uncertainty which is given by $\Delta X \Delta P > h$, leads to a competition between the potential energy and the kinetic energy and the physics is determined by the ratio of both. When $E_C/E_J < 0.001$, the system is in an ordered vortex lattice state, and when the rotational frequency $\Omega$ becomes larger than a critical value, $E_C/E_J \geq 0.001$ and the system undergoes a QPT and enters a melted vortex state.

In conclusion, according to the above analysis, we see that the Abrikosov-Tkachenko vortex lattice may be equivalently being understood as an array of Josephson junctions. The Tkachenko modes are basically equivalent to the low energy excitations (Goldstone modes) of an ordered phase. For large vortex densities $n_V$, the lattice is expected to melt, at zero temperature due to quantum fluctuations and the vortices form a highly correlated quantum liquid. The vortex melting is a result of quantum fluctuations around the ordered state due to the low energy excitations (Tkachenko modes). We show that a vortex liquid phase appears for $E_C/E_J \geq 0.001$.

**Acknowledgements:** I acknowledge support by the Abdus Salam International Centre for Theoretical Physics, Trieste, Italy under the ICTP-TRIL fellowship scheme. My thanks to Prof. E. Arimondo, for providing me the facilities for this work at INFM, Pisa. Also I am grateful to T. K. Ghosh for useful discussions.

FIGURE CAPTIONS

Figure 1: The Tkachenko mode frequencies as a function of $\Omega/\omega_\rho$ for $N=2.5\times10^6$, $\{\omega_z,\omega_\rho\}=2\pi\{8.3,5.2\}$Hz. Also shown are the data points.

Figure 2: The filling factor $\nu$ vs. $\Omega/\omega_\rho$ for $N=2.5\times10^6$, $\{\omega_z,\omega_\rho\}=2\pi\{8.3,5.2\}$Hz. For rapid rotations of about $\Omega=0.975\omega_\rho$ (maximum rotation achieve until now), melting of the vortex lattice is expected to occur at $\nu\approx17$.

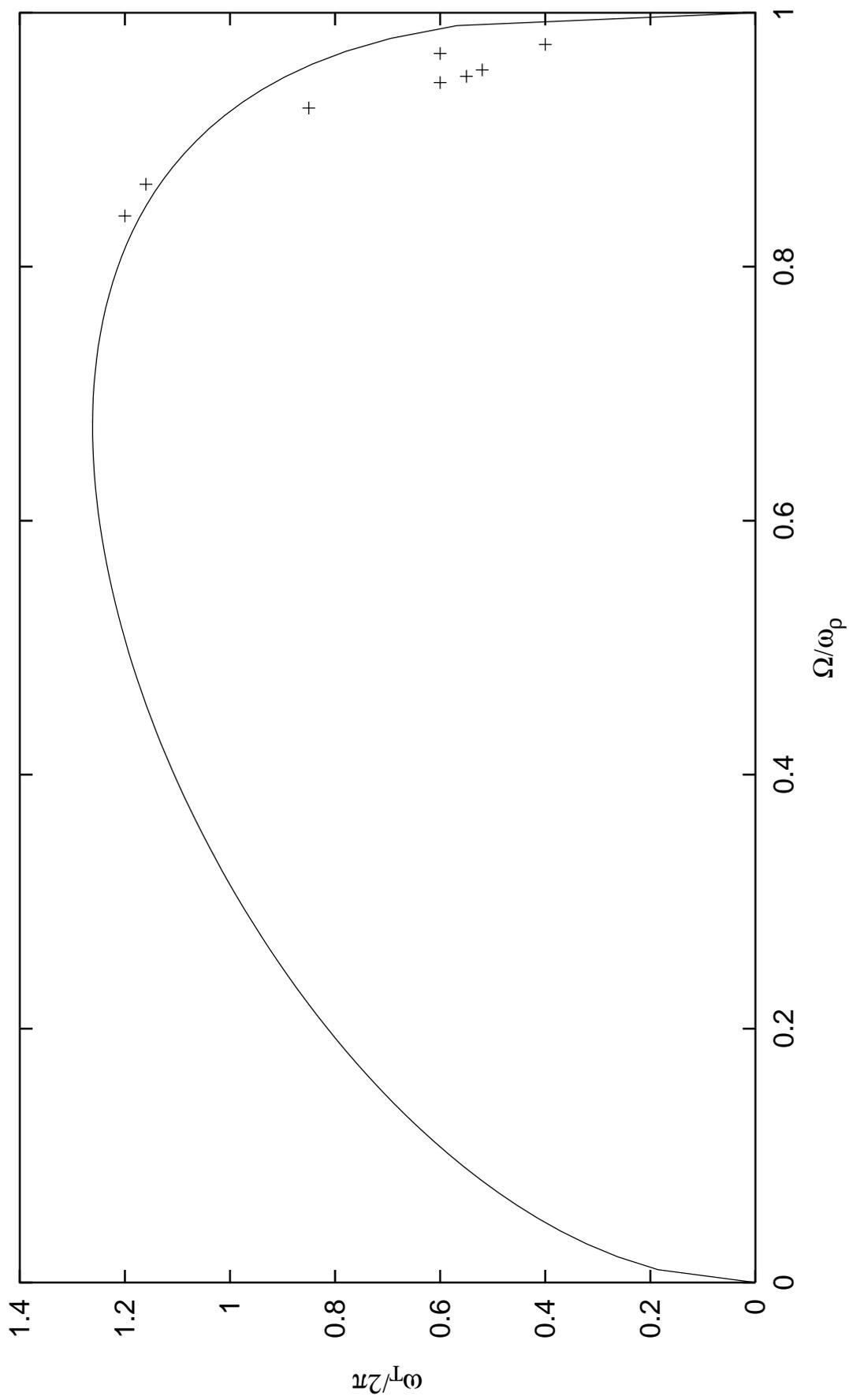

Figure 1

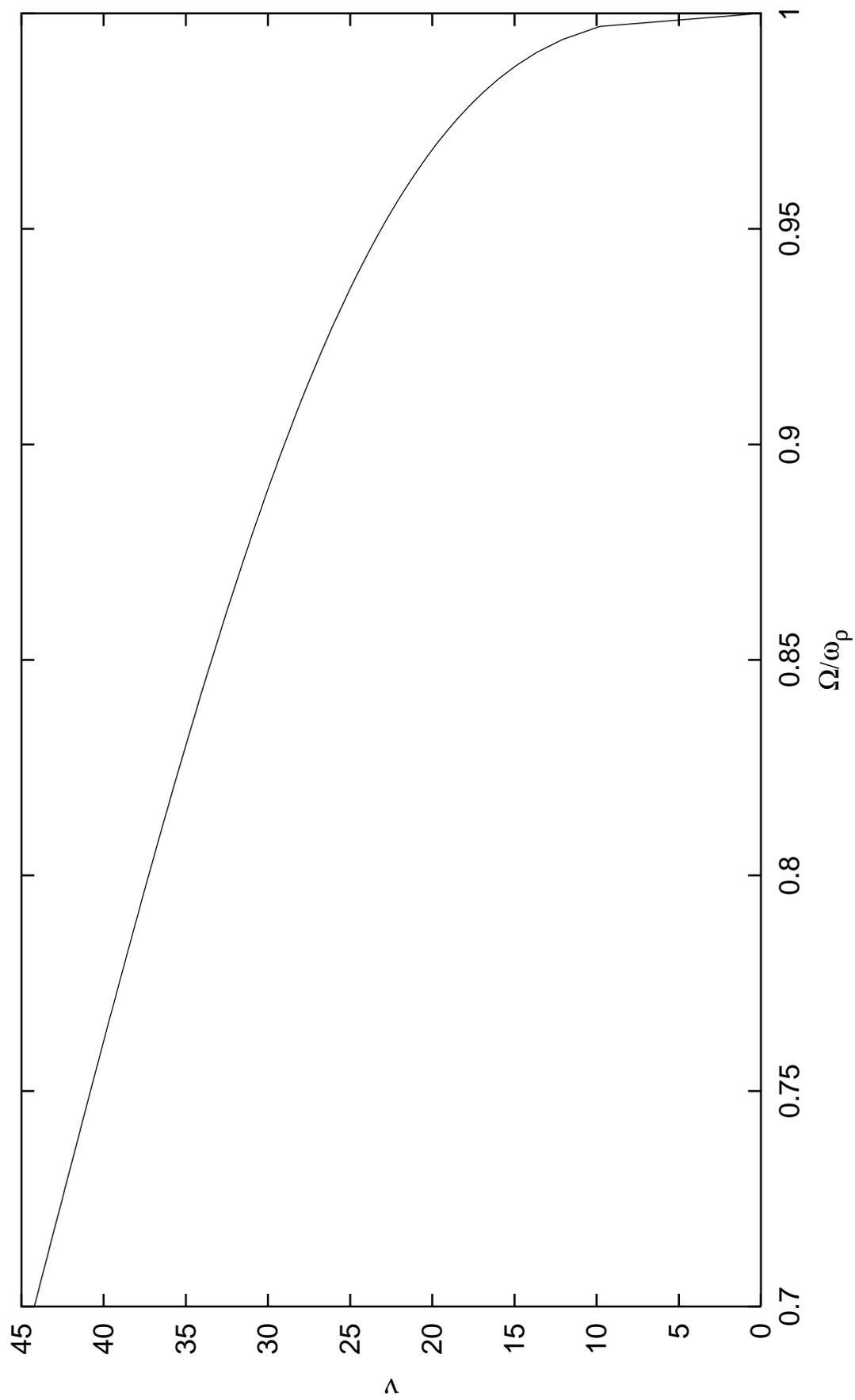

Figure 2